\documentclass[twocolumn]{pasj01}

\Received{$\langle$reception date$\rangle$}
\Accepted{$\langle$acception date$\rangle$}
\Published{$\langle$publication date$\rangle$}
\newcommand{\hii}{H~II~}
\newcommand{\hiirs}{\hii~regions}
\usepackage{soul}
\usepackage{color}

\newcommand{\uchii}{UC~\hii}
\newcommand{\uchiir}{UC~\hii region~}
\newcommand{\uchiirs}{UC~\hii regions~}

\newcommand{\cuna}{~}
\usepackage{appendix}
\usepackage{natbib}
\setcitestyle{authoryear,open={(},close={)}}
\usepackage{lineno}
\usepackage[T1]{fontenc}

\usepackage{savesym}
\savesymbol{iint}
\savesymbol{leftroot}
\savesymbol{uproot}
\savesymbol{iiint}
\savesymbol{idotsint}
\savesymbol{iiiint}

\usepackage{amsmath}
\usepackage{amstext}

\usepackage{txfonts}

\restoresymbol{AMS}{idotsint}
\restoresymbol{AMS}{iint}
\restoresymbol{AMS}{leftroot}
\restoresymbol{AMS}{uproot}
\restoresymbol{AMS}{iiint}
\restoresymbol{AMS}{iiiint}

\begin{document} 

\Received{2022/07/15}
\Accepted{2022/11/3}
\Published{2022/11/30}

\title{The Ultracompact regions G40.54+2.59 and G34.13+0.47: A New Detection of Compact Radio Sources\thanks{Thesis submitted by Harold Viveros as a partial fulfillment for the requirements of Ms. Sc Degree in Astronomy, Universidad de Guanajuato}}

%%% begin:list of authors
% Do NOT capitalize all letters in "textsc".

\author{Harold E. \textsc{Viveros}\altaffilmark{1}}
\altaffiltext{1}{Departamento de Astronom\'{i}a, Universidad de Guanajuato, Apartado Postal 144, 36000, Guanajuato, Guanajuato, M\'exico}
\email{harold@astro.ugto.mx}

\author{Josep M. \textsc{Masque}\altaffilmark{1}}
%\altaffiltext{1}{Departamento de Astronom\'{i}a, Universidad de Guanajuato, Apartado Postal 144, 36000, Guanajuato, Guanajuato, M\'exico}
\email{masque@astro.ugto.mx}

\author{Miguel A. \textsc{Trinidad}\altaffilmark{1}}
%\altaffiltext{1}{Departamento de Astronom\'{i}a, Universidad de Guanajuato, Apartado Postal 144, 36000, Guanajuato, Guanajuato, M\'exico}
\email{trinidad@astro.ugto.mx}

\author{Eduardo \textsc{de la Fuente}\altaffilmark{2,3}}
\altaffiltext{2}{Departamento de F\'{i}sica, CUCEI, Universidad de Guadalajara, Blvd. Marcelino Garc\'{i}a Barragan 1420, Ol\'{i}mpica, 44430, Guadalajara, Jalisco, M\'exico}
\email{eduardo.delafuente@academicos.udg.mx}
\altaffiltext{3}{Institute for Cosmic Ray Research (ICRR), University of Tokyo, 1--5 Kashiwanoha 5-Chome, Kashiwa, Chiba 277-8582, Japan (Sabbatical 2021)}

\KeyWords{HII regions -- stars: formation -- stars: massive -- radio continuum: ISM -- ISM: individual objects (G40.54+2.59, G34.13+0.47)}

\maketitle
%\draft

\begin{abstract}

We report the detection of three compact ($< 0.001$ pc) radio sources (CRSs) at K$_{a}$-band (0.9 cm) in the \uchiirs G040.54+2.59 (two CRSs) and G034.13+0.47 (one CRS). These CRSs have weak flux densities and are located at the center of their respective \uchii regions. We found no clear association between massive ionizing stars and CRSs but some radiative influence on the latter, as suggested by their large emission measures (> $10^7 \mathrm{cm}^{-6}\mathrm{pc}$), typical of photo evaporating neutral objects close to or associated with massive stars. Our modelling of G40.54+2.59 shows that their CRSs supply enough ionized material to shape its morphology while significantly extending its observable lifetime. On the other hand, despite the possible relation of the CRS with the large-scale outflow signatures observed in G034.13+0.47, the influence of this CRS on the evolution of the \uchii region is unlikely. Our results show that the presence of CRSs can alleviate the so-called lifetime problem of UCHII regions. Still, to address their dynamical evolution adequately, the scenario must include additional mechanisms like ambient confinement, or the role of the kinematics of their associated stellar objects.

\end{abstract}

%\linenumbers

\section{Introduction}
\label{sec:intro}

When powerful Lyman continuum emission is produced by very young massive OB stars still embedded in their natal molecular cloud, the so-called {Ultra-compact HII} regions ~\citep[UCHII,][]{harris1973,kurtz2002} grow around these stars. 
These regions are bright mainly at centimeter radio and infrared wavelengths (peaking at $\sim100\ \mu$m), and are characterized by their small sizes, $d\leq 0.1$ pc, high electron density, $n_{e}\geq 10^{4}$ cm$^{-3}$, and high emission measure\footnote{The optical depth, $\tau$, is parametrized commonly through EM, so that it is customary to refer to HII regions according to their emission measure as well.}, EM $\geq10^{6}$ pc cm$^{-6}$~\citep{habing-israel1979, 2002ARA&A..40...27C, 2020MNRAS.497.4436D}. Other kind of compact HII regions, with typical sizes and densities of $d\leq 0.03$ pc and $n_{e}\geq10^{6}$ cm$^{-3}$, respectively, are called {Hyper-Compact} HII regions \cuna\citep[HCHII,][]{2005IAUS..227..111K, 2011ApJ...739L...9S}. Despite their extremely compact size, the canonical classification of the youngest HII regions based on the linear size should be treated with special care~\citep{yang2019}. In any case, HCHII regions constitute the most compact HII regions observed so far and possibly harbor a single star or binary system. 

From the VLA survey of~\citet{wood-churchwell1989} on 75 UCH II regions, two essential aspects of the nature of the UCHII regions have been revealed. First, there exist five different morphologies associated with their physical properties: cometary, core-halo, shell, irregular or multi-peaked, and bipolar~\citep[later introduced by][]{depree2005}. The second aspect is related to their dynamical time scales. Specifically, the presence of neutral and ionized gas in the boundary of the UCHII regions should exert a high-pressure contrast causing a hydrodynamic non-equilibrium expansion of the ionized material. Consequently, these regions must grow at speeds equal to or greater than the sound speed ($\sim$10 km s$^{-1}$), so that the UCHII regions would not be detected after $\sim 10^{4}$ yr (e.g., 1.7\% of the lifetime of an O6 star). However, the $\sim$10\% - 20\% of all the O stars surveyed by \citep{wood-churchwell1989} is associated with UCHII regions, much more than expected theoretically by simple hydrodynamical arguments. The latter implies that a successful theory on UC HII region evolution should address their morphology and expected lifetime.

To explain the morphological structure of the \uchii regions, promising models aimed principally at explaining the cometary one surges. One of these models is the so-called ``champagne flow'' or ``blister model''~\citep{tenorio-tagle1979, yorke1983}. In this model, most massive stars form near the boundaries of dense inter-clouds. When the HII region grows around these stars, it ``breaks out'' the cloud-inter cloud border, and a ``blowout'' of photo-evaporated material is driven due to overpressure, giving the appearance of a cometary morphology. Likewise, in-homogeneities in the ambient density of the massive stars have also been appealing to explain this type of morphology~\citep{fey1995}. The drawback to these models is that they do not support the additional factor of 10 to 100 in the scale of the dynamical age observed in the \uchiirs and the ``cometary'' class represents only about 20\% of these regions ~\citep{wood-churchwell1989}. In a most comprehensive scenario that accounts for both morphology and dynamical age, the motion of the ionizing stars relative to their natal molecular cloud creates a bow-shocked emission supported by the wind of the stars itself and the inward ram pressure of the ambient gas. At this stage, the cometary morphology would be featured when viewed ``from the side'', and the core-halo morphology when viewed ``head-on'' (or ``tail-on'')~\citep{vanburenetal1990, mac-low1991, garay1994}. If the motion of stars is almost negligible concerning the molecular cloud, a spherical or shell-like signature appears. Despite the completeness of this model, some authors have found observational inconsistencies on its predicted velocity structure~\citep{lumsden-hoare1996}. Possibly, extra elements such as a non-homogenous medium should be included in the model, but, in any case, it provides a first approach to explain the variety of morphologies observed for UCHII regions.

Another feasible scenario aimed at solving the long lifetime of \uchii regions (regardless of their morphology) considers a certain mechanism of continuous ``replenishment'' of ionized material into the region. For instance, an ionizing massive young star with a circumstellar disk which is being continuously photo-evaporated by its own UV emission, emanating material into the \uchii region~\citep{hollenbach1994, lugo2004}. Likewise,~\citet{lizano1996} and~\citet{johnstone1998} assume that circumstellar disks or dense clumps of gas present in the region are being continuously photo-evaporated by the UV emission of external massive stars. In a dynamically opposite scenario, ~\citet{peters2010} and~\citet{De-pree2014} suggest that the accretion flow onto the massive stars transfers material so quickly to the HII region, preventing its development through the cloud. An exciting possibility was raised by~\citet{keto2006, keto2007} studying the progression of HII regions, including rotation. They found that this evolution goes from a quasi-spherical gravitationally trapped ionized region (resembling an HCHII region at initial stages) to one with a bipolar morphology. In the equatorial plane of the bipolar morphology, a photo-evaporating disk is formed. Thus, this possibility suggests a unified paradigm between ``infalling'' and ``replenishing'' models.
 
From an observational point of view, interferometric radio observations are crucial to studying the \uchii regions. Their morphology and internal structure can reveal important clues to their evolution, which is reached only when the arrays operate at their most extended configurations. In this sense, the presence of a rich variety of Compact Radio Sources (CRSs) of enigmatic nature associated with \uchii regions has been recently revealed~\citep{kawamuramasson1998, carral2002, rodriguez2014, medina2018}. At first, glance, identifying these CRS as HCHII is tempting given their similar compact size (a few hundreds of AU). However, a deeper exploration of the CRS properties points to a different nature~\citep{masque2021}. For example,~\citet{dzib2013} found free-free emission in the UCHII region W3 (OH) originating from a CRS. It is thought that the emission may originate from the debris of a photo-evaporating disk.

On the other hand, non-thermal emission produced by a magnetically active star has been found in the Monoceros region R2, towards an \uchii region known previously~\citep{2016ApJ...826..201D}. All these results show the wealth inherent to the star-forming regions, whose multiplicity is possibly manifested from early phases. Recently,~\citet{masque2017} have found 13 CRS associated with 9 UCHII, suggesting that all UC HII regions have at least one CRS associated. Possibly, their weak emission (a few mJy as much) has been hidden by the bright ionized gas of the UC HII region, and most of them only become 'visible' nowadays with the new pioneering instrumental capabilities. This fact aroused a strong incentive for explaining the nature of the CRSs because they could be the missing link in understanding the physical process occurring in the UC HII regions.

%and could be in accordance with the ``replenishing'' models.

In this paper, we report by the first time the detection of CRSs in the {G34.13+0.47} (G34.13) and  {G40.54+2.59} (G40.54) \uchiirs using deep radio observations, as well as discuss the relation of the possible nature of these CRSs with the dynamical evolution and morphology of their host \uchiirs. The regions were chosen because they satisfy the following two criteria: (i) they are located at less than 2-3 kpc far away; (ii) they have well-defined extended structure with large emission measures suggesting that, if the ionized emission of the \uchii region comes from photo-evaporated material, these regions must contain at least one CRS associated to them~\citep{2017A&A...598A.136M}.

In the next paragraph, we shall have an overview of the regions under study. In {Section}.~2 we depict our radio observations and the filtering method to suppress the extended emission. In {Section}.~3, a detailed description of the CRSs and their derived physical features is made, as well as a near-infrared and morphological analysis of the regions using the 2MASS Point Source and UKIDSS-GPS catalogs, and hydro-dynamical models, respectively. In {Section}.~4 we discuss the results in the context of the evolution of the \uchiirs. Finally, in {Section}.~5, we enumerate the conclusions of our study.

\emph{G34.13 and G40.54} These regions were previously observed with the VLA at 5 GHz in B-configuration as part of the RMS and CORNISH-projects~\citep{urquhart2009,purcell2013}. At this frequency, both \uchiirs have angular sizes of $\sim9''$, and integrated flux densities of $\sim$ 450 mJy, which implies a measure emission of EM $\sim 5\times 10^{5}\ \mathrm{cm}^{-6} \mathrm{pc}$~\citep{1969ApJ...156..269S}, at a distance of $\sim$2.3 kpc to this region~\citep{wienen2015, reid2016}. Their radio flux density implies that these regions must be excited by a B0.5V star by assuming that the rate of ionizing photons is equal to that of the hydrogen recombinations in these regions\cuna\citep{2003ApJ...599.1333S}. This stellar classification is also consistent with their bolometric luminosities ($\approx 1.5\times 10^{4}$ L$_{\odot}$)~\citep{2011A&A...525A.149M, 2017MNRAS.471..100E}.

%%%%%%%%%%%%%%%%%%%%%%%%%%%%%%%%%%%%%%%%%%%%%%%%%%%%%%%%%

\section{Observations and data reduction}
\label{sec:observations}

The Karl Jansky Very Large Array (JVLA) observations at K$_{a}$ band (0.9 cm)  were conducted on 2017 December 28 with the B configuration toward the \uchiirs G034.13+0.47 and G040.54+2.59, with the phase centers of our observations being at J2000 coordinates of $18^\mathrm{h}51^\mathrm{m}57^\mathrm{s}24$  \& $-01^{\circ}17'02\rlap{$''$}.5$, and $18^\mathrm{h}56^\mathrm{m}04^\mathrm{s}56$  \& $07^{\circ}57'28\rlap{$''$}.6$, respectively. These regions are found relatively close in the sky ($\leq$ 10 deg), so the same phase calibrator, J1851+0035, was used for both of them. One initial pointing was performed on this calibrator, and the derived solutions were applied to the rest of the scans. The absolute flux density calibrator was 3C286, for which bandpass calibration solutions were also obtained. Our observations spanned a total time of $\sim$90 min, with a total time on-target of $\sim$20 min per region. We established the three-bit set samplers in full polarization mode and 3 s as integration time. The central frequency was 33 GHz with basebands centers at 30, 32, 34, and 36 GHz, giving a total bandwidth of 8.0 GHz (see {Table}.~\ref{tab:obs_outline} for a sketch of our observations).

To edit, calibrate and image the data, we used the Common Astronomy Software Applications (CASA)\footnote{https://casa.nrao.edu/} from the NRAO. To distinguish weak compact emission and to assess that it is genuine and not spurious emission resulting from residual extended emission of the \uchii region, we CLEANed the calibrated data several times, employing different ranges of visibilities. The largest range is the result of including only base-lines greater than 111 k$\lambda$ (this is equivalent to removing angular scales larger than 0.9 arcsec), while the shortest range corresponds to including only visibilities > 444 k$\lambda$ (equivalent to removing angular scales larger than 0.5 arcsec). In between these largest and shortest ranges, we explored additional $uv$-ranges moving the lower limiting visibilities in steps of $\delta$($uv$) = 50 k$\lambda$. This results in a set of maps for each \uchii region that includes all the explored $uv$-ranges. Once we identified a compact feature in the regions, we compared its integrated flux density between maps corresponding to different $uv$-ranges and considered it a real structure (i.e., CRS) of no variations in its flux density measured between maps were appreciable. These criteria provide a final map with the adequate $uv$-range to properly isolate the CRS from extended emission. The choice of $uv$-ranges is based on the CRS sizes reported in \citet{masque2017} and sweeps a sufficiently large number of mapped angular scales for appropriate removal of extended emission. So, for one first set of maps, a WEIGHTING = 'NATURAL' was chosen to gain sensitivity. A compact source was found in G34.13+0.47 region, and two others in G40.54+2.59. We have named these sources G34.13-VLA1, G40.54-VLA1 and G40.54-VLA2, respectively. 

In the second set, we searched for structure in these compact sources, so new maps were constructed by setting WEIGHTING  = ``Briggs'', with ``robust'' = 0.0. As we want the best compromise between angular resolution and sensibility, the map with suppressed base-lines shorter than 278 k$\lambda$ was selected for both weightings. The resulting maps achieve rms noises and synthesized beams of $\sim$17 $\mu$Jy beam$^{-1}$ and $\sim$0.2$''$, respectively.

\section{Results and Analysis}
\label{sec:observ-results}

In this section, we describe and inspect the properties of the three CRSs found in the \uchiirs G34.13 and G40.54, and reported here for the first time. Gaussian fittings were made to extract the peak and integrated flux density, peak position and angular sizes for the CRSs as reported in {Tables}~\ref{tab:obs-results}. Only for G40.54-VLA1 the beam-deconvolution method converges. For the remaining unresolved sources, upper limits of half of the beam size are reported (Column 4). Physical parameters were derived from the flux density and angular sizes reported in the {Table}.~\ref{tab:obs-results} by following standard expressions (see appendix A). Our 0.9 cm continuum maps resulting from the removal of extended emission (see {Section} \ref{sec:observations}) compared to the 6 cm emission presented in~\citet{urquhart2009} is shown in {Figures}.~\ref{fig:34.13_ourcontourRFCsuper} and~\ref{fig:G40.54_ourcontourRFCsuper} (blue contours vs. black contours and color scale, respectively).

\subsection{G34.13 Region}

\subsubsection{Radio Continuum}
\label{sec:Results_G3413_radiocontinuum}

%In {Figure}~\ref{fig:34.13_ourcontourRFCsuper} we superimposed the 6 cm emission with that of 0.9 cm from our observations for G34.13. 

\begin{figure*}
\begin{center}
	\includegraphics[width=0.45\textwidth]{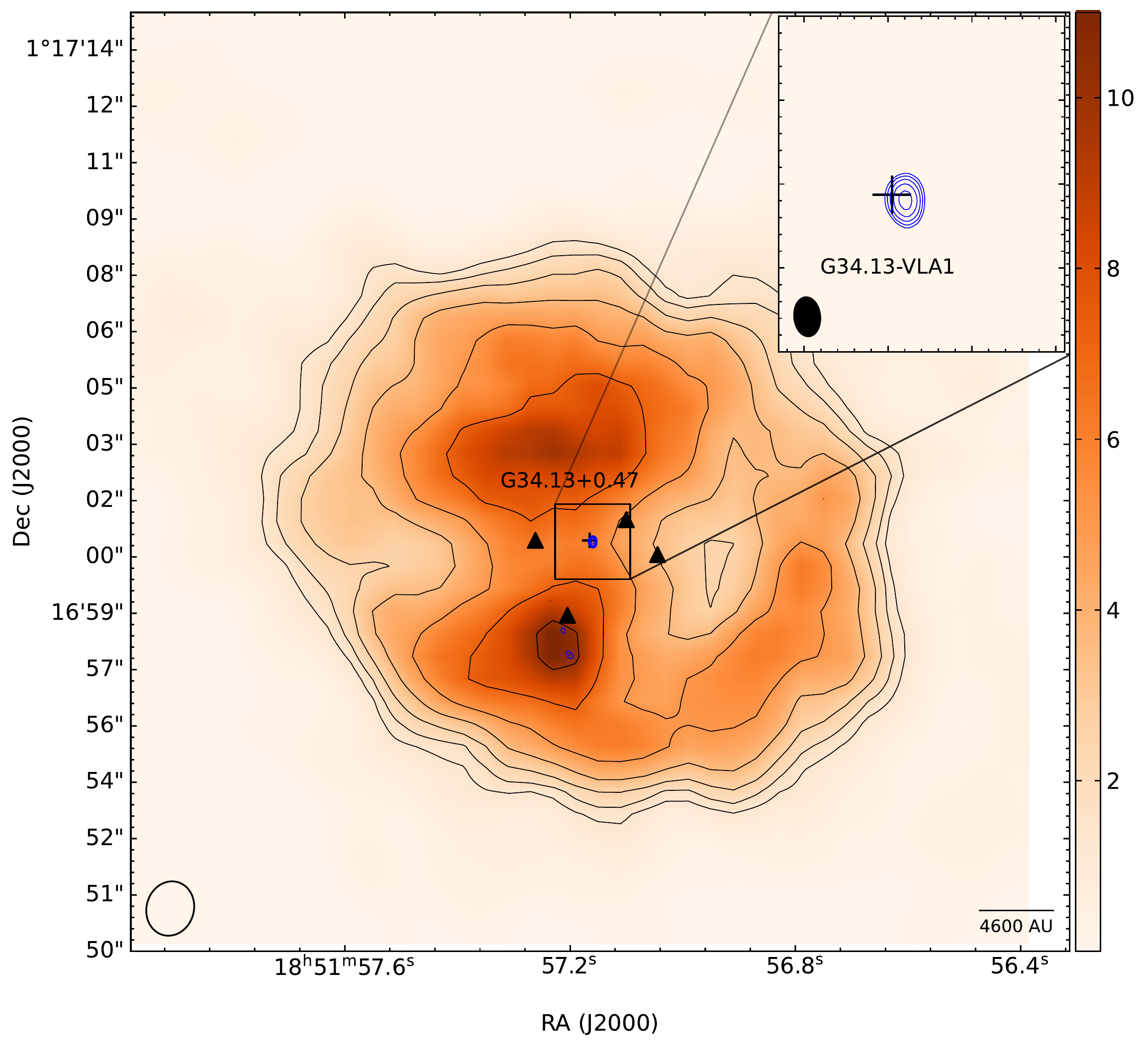}
	\includegraphics[width=0.45\textwidth]{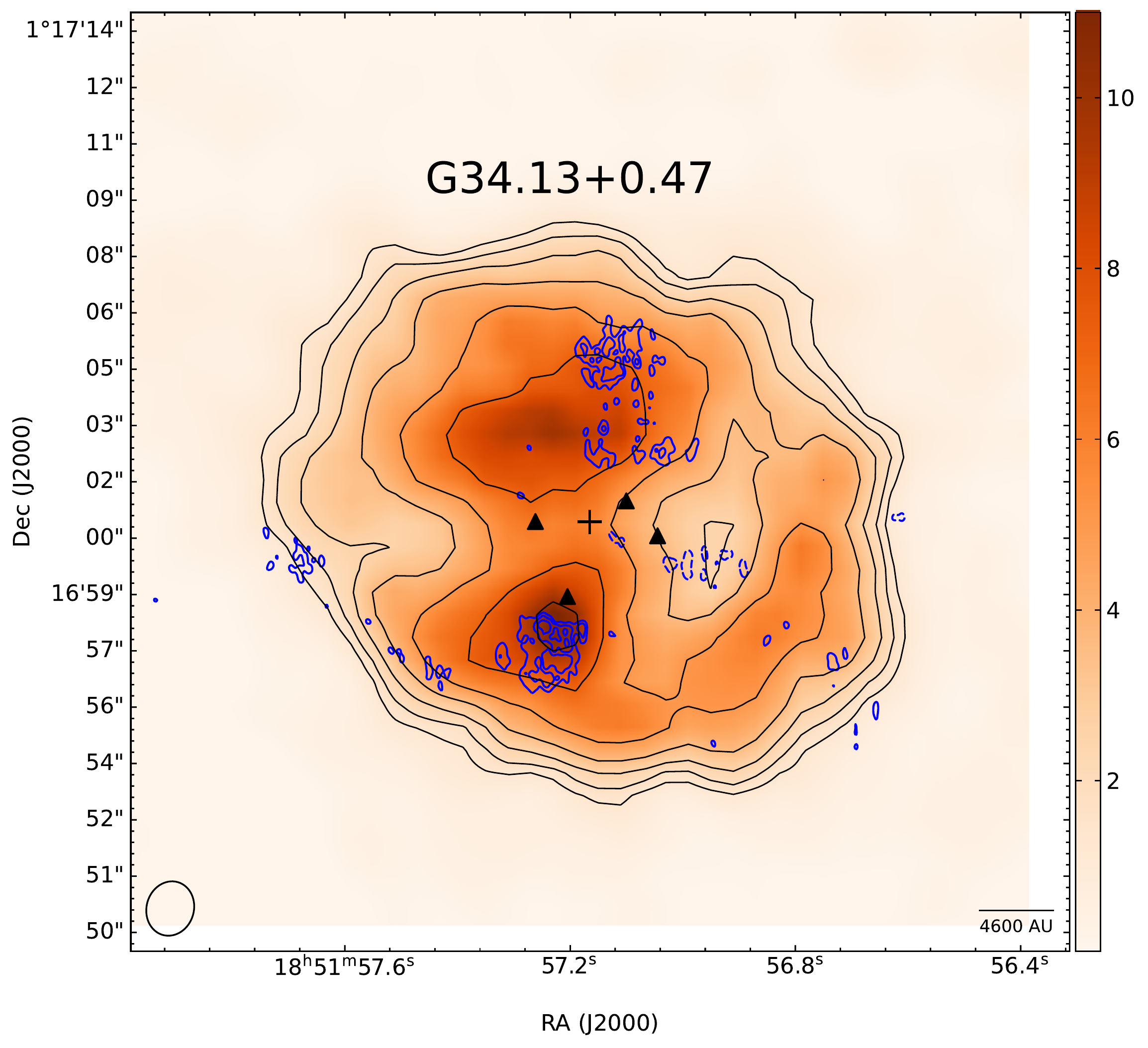}
\end{center}
\caption{Left: Superimposed contours at 6 cm (black lines and color scale) from~\citet{urquhart2009} with our contour maps at 0.9 cm with short spaced visibilities removed (see text, blue contours) for the \uchiir  G34.13. The 6 cm contours are N times -2$^0$ to 2$^{0}$, 2$^{\mathrm{1/2}}$, 2$^{1}$, 2$^{3/2}$, 2$^{2}$, 2$^{5/2}$, 2$^{3}$, 2$^{7/2}$, 2$^{4}$, where N is 5 times of the $rms$ noise of the map (0.25 mJy beam$^{-1}$). The same level spacing is used for the 0.9 cm emission but with the corresponding noise level of the map (15 $\mu$Jy beam$^{-1}$). In the upper right panel, we show a ``zooming-in'' of the CRS with the corresponding 0.9 cm beam (fill oval at the lower-left corner). The ``plus-symbol'' over G34.13-VLA1 denotes the near-infrared UKIDSS-GPS K-band detection. Filled black triangles are the UKIDSS-GPS point source near G34.13-VLA1 ($\leqslant$ 2.5 arcsec) and are only detected in the K-band. The synthesized beam sizes for 6 and 0.9 cm are shown in the bottom left corner of large and small panels, respectively. Right: Same as Left, but considering all baselines (blue contours). In this case, the \textit{rms} noise of the 0.9 cm map is $\approx$ 15.0 $\mu$Jy beam$^{-1}$, and the contour levels follow the same sequence as for the 6 cm map but with N=6. The synthesized beam size of the 0.9 map is $0.30'' \times 0.20''$  (P.A. 3.0$^{\circ}$)}   

\label{fig:34.13_ourcontourRFCsuper}
\end{figure*}

\begin{figure*}
\begin{center}
	\includegraphics[width=0.45\textwidth]{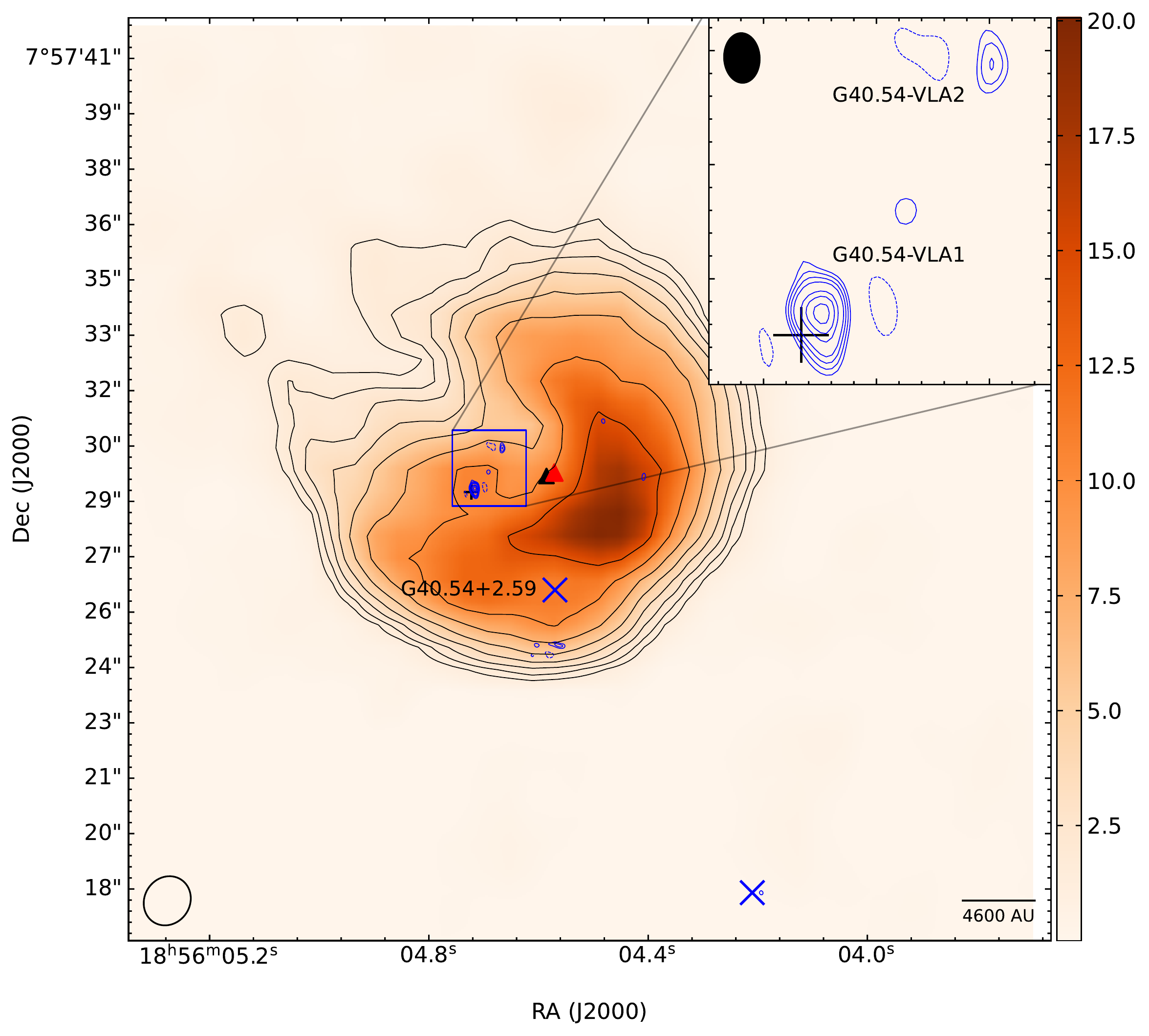}
    \includegraphics[width=0.45\textwidth]{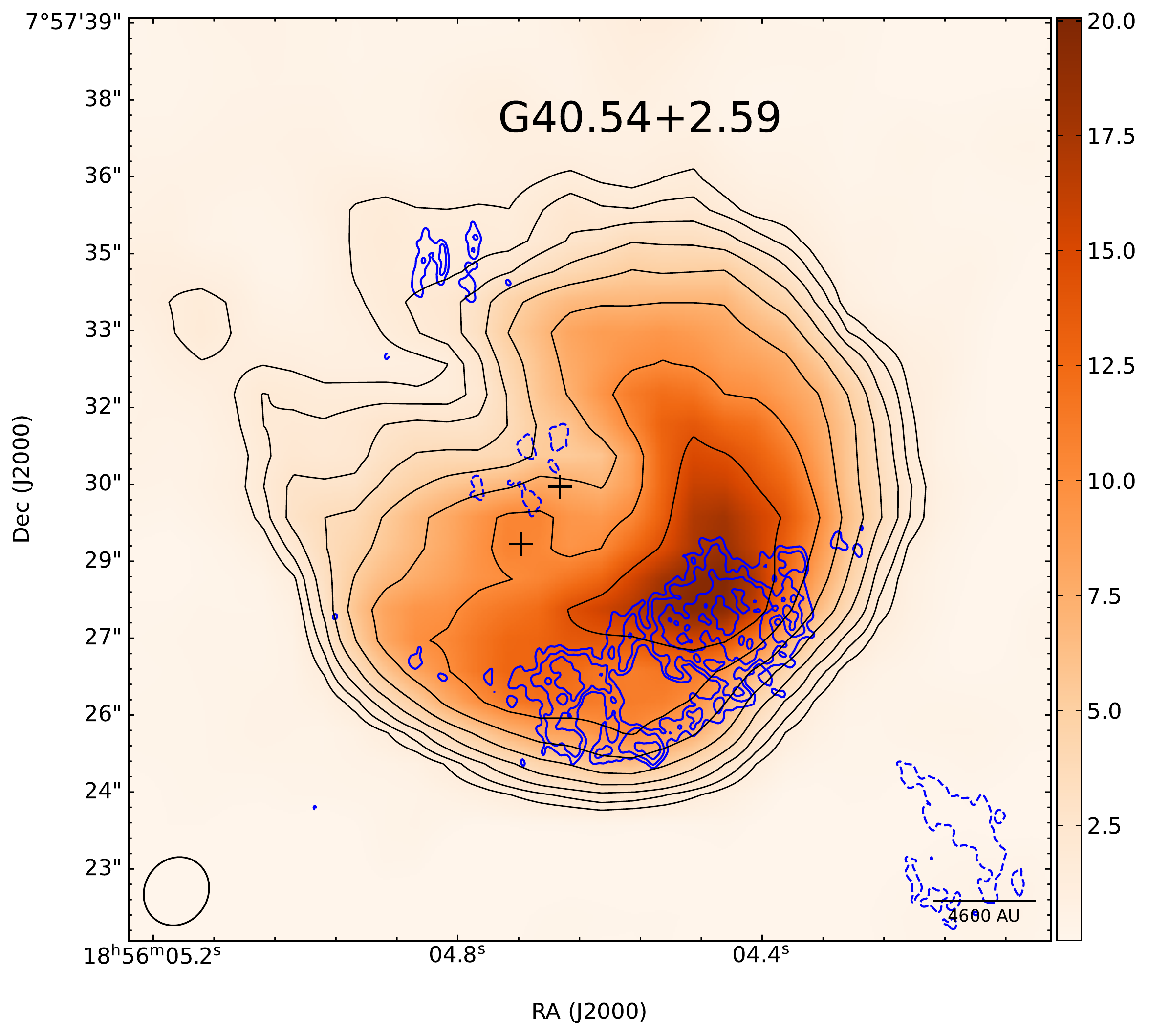}
\end{center}
\caption{Left: Same as Figure~\ref{fig:34.13_ourcontourRFCsuper} but with the 0.9 cm contours -3.0, 3.0, 5.0, 6.0, 6.5, 7.0, 7.5, and 8.0 times the rms noise of 16.5 $\mu$Jy beam$^{-1}$ of the map for the \uchiir G40.54. The empty and full oval in the left corner of the images represent the synthesized beam of 6 and 0.9 cm maps, respectively. Marks as a ``cross'' denote the infrared sources detected by~\citet{ramirez-alegria2014}. The red and black-filled triangles mark the position of the detected 2MASS and UKIDSS infrared point sources, respectively. Right: Same as Left but considering all baselines (blue contours). In this case, the \textit{rms} noise of the 0.9 cm map is 18.0 $\mu$Jy beam$^{-1}$, and the contour levels follow the same sequence as for the 6 cm map but with N=6. The synthesized beam size of the 0.9 cm map is $0.28'' \times 0.20''$  (P.A. 0.5$^{\circ}$)}

\label{fig:G40.54_ourcontourRFCsuper}
\end{figure*}

The only detected CRS in this region, G34.13-VLA1, has been highlighted in the upper right panel of Figure~\ref{fig:34.13_ourcontourRFCsuper}, in which we have used a cross symbol to denote UKIDSS-GPS near-infrared K-band emission associated. As can be seen, the CRS is highly centered in the shell-like structure of the 6 cm emission, suggesting that G34.13-VLA1 can be associated with the ionizing source of the \uchiir~\citep[e.g.,][]{carral2002}. Furthermore, its location between the two bright extended lobes nearly in the north-south direction observed in the 6 cm map makes G34.13-VLA1 presumably connected with some expansion process occurring in the region~\citep[e.g.,][]{tafoya2004}.

The first possibility to explore in line with the geometrical situation described above is that the CRS harbors the star that excites the region and drives its dynamics. In this case, all the Lyman photons produced by the star can contribute to ionizing the CRS material. Numerically, under the Str\"omgren formalism for ionized gas in equilibrium and assuming that each photon ionizes one hydrogen atom, the corresponding minimum photon rate required for ``switching on'' G34.13-VLA1 is $N_{\mathrm{CRS}} = 10^{44} $ s$^{-1}$ (seventh column in {Table}.~\ref{tab:physical-param}). According to~\citet{panagia1973}, this would correspond to, at least, either a spectral type B3III or B2V star. Besides, such a star B3 delivers the expected flux density at UKIDSS near-IR K-band corrected for extinction (see Viveros et al. in prep.). The agreement between spectral types of the stellar object in G34.13-VLA1 derived independently from IR and radio data suggests a minimum UV photon leakage/loss. Otherwise, an earlier spectral type would be derived from the IR data. At the same time, since
the radio continuum emission indicates that at least a B0 star is required to excite the overall G34.13 region, the above agreement implies no direct association of such massive star on G34.13-VLA1. 

Other IR sources seen in the map (see Fig. \ref{fig:34.13_ourcontourRFCsuper}) could be associated with this massive star, but their available photometric information is insufficient to determine their spectral type accurately. 
However, their close position to G34.13-VLA1 seen in the map raises the possibility of some external radiative influence on the CRS by some of them. To quantify somehow the amount of ionization caused by the four UKIDSS-GPS sources, we considered the contribution of the nearby IR source (1.5 arcsec of distance) assuming a spectral type of B0. Under the scenario of external photoevaporation, the ionizing photon rate impinging the CRS, $N_{A}$, should fulfil the rate needed to excite the CRSs, $N_{\mathrm{CRS}}$. The former rate is the result of the photon flux dilution by the distance of the CRS from the star producing the photons and, taking into account the area of the CRS exposed to the radiation, it can be estimated with the following expression:

\begin{equation}\label{eq:dilution}
    N_{A} = \frac{\theta_{\mathrm{eff}}^2}{16}\frac{N_{*}}{d^{2}},
\end{equation}\

where $\theta_{\mathrm{eff}}\equiv\sqrt{\theta_{\mathrm{maj}}\theta_{\mathrm{min}}}$ is the equivalent angular size of the CRS calculated from the deconvolved results shown in Table \ref{tab:obs-results}, $d$ is the angular separation between the CRS and the exciting star, and $N_{*}$ is the Lyman photon rate delivered by the latter (i.e. $10^{48}$ s$^{-1}$ for a B0 star). In\cuna{Table}\cuna\ref{tab:distance_star_CRS} we compare the values of $N_{A}$ and $N_{\mathrm{CRS}}$. This calculation assumes that all the objects are in the plane of the sky, and the reported angular distances are upper limits. In addition, the deconvolved sizes of the CRS are upper limits or very uncertain. The reported values in Table\cuna\ref{tab:distance_star_CRS} for G34.13-VLA1 enables some external influence by the IR sources on this CRS. We conclude that, although the B3 star embedded in G34.13-VLA1 provides an important contribution to ionize the CRS, we cannot discard some external influence of the IR sources detected in the region.

%Finally, we added each individual contribution of the four IR sources, which also yields an upper limit for the overall contribution. 

\subsubsection{The connection between G34.13-VLA1 and the large-scale outflow signatures seen in the region}

Mid-infrared observations with the Galactic Legacy Mid- Plane Survey Extraordinaire \citep[GLIMPSE][]{2005ApJ...630L.149B} reveal a large-scale bipolar structure ($\sim$ 90 arcsec) in the Southeast-Northwest direction, probably being an outflow powered by a Class I Young Stellar Object (YSO) as discussed by \citet{2009MNRAS.393..354P}. Furthermore, these authors inferred the presence of a dusty toroid, of 10.2 arcsec, centered on the bipolar structure of this source. In this scenario, this structure could be tracing the part of the ionization front of the UCHII region of G34.13 interacting with the toroid \citep[e.g.,][]{torrelles1985,rodriguez1986}

The 8 GHz bandwidth of our observations and the relatively high signal-to-noise ratio of G34.13-VLA1 enables us to determine its spectral index. By applying the general relation: $\alpha = \log (S_{\nu_{1}}/ S_{\nu_{2}}) / \log (\nu_{1}/\nu_{2}$), we calculated a value of $\alpha$ of $0.93\pm 0.38$ corresponding to partially thick free-free emission. Within the uncertainty, this spectral index is consistent with the presence of an ionized stellar wind, with the electron density following a radial gradient~\citep[$\alpha=0.6$,][]{panagia-felli1975}. This result is independent of the geometry of the wind, meaning that a collimated one (e.g. ionized protostellar jet) gives the same index \citep{1986ApJ...304..713R,2018A&ARv..26....3A}. 

All the elements discussed in this section raise the question of whether the free-free emission of G34.13-VLA1 comes from a possible unresolved protostellar radio jet that powers the large-scale outflow or not. To test this possibility we compared the lower limit for EM of G34.13-VLA1 ($\approx 4\times 10^{7}$ cm$^{-6}$pc, see Table~\ref{tab:physical-param}), with those values typically expected for radio jets ($\sim10^{3} \mathrm{cm}^{-6}\mathrm{pc}$,~\citealt{rodriguez1990, curiel1993}). Even considering the harsh radiative ambient of the \uchiir from where the putative jet is exposed, a resulting fully ionized jet would not provide sufficiently ionized material to match the EM reported in the table, given the typical ionization fraction inferred for jets found in other more quiescent environments is very low \citep[$\leq 14\%$, see e.g][and references there in]{2018A&A...612A.103C}. Thus, the contribution of other mechanisms producing ionized gas additional to an isolated jet must be present in the CRS and become dominant in our observed free-free emission in G34.13-VLA1. 

%~\citep[$\sim10^{3} \mathrm{cm}^{-6}\mathrm{pc}$,][]{rodriguez1990, curiel1993}

In this sense, other theoretical scenarios in line with those argued in previous section predict EM consistent with those derived by us, namely photo-evaporating disk winds or trapped HCHII regions~\citep[$\sim$10$^8$ cm$^{-6}$ pc,][]{hollenbach1994, lugo2004, keto2006, keto2007}. These objects are expected to have small sizes well below our beam size: as an example, for the particular case of a B3 star, the gravitational radius where the ionized gas can be trapped becomes $r_{g} = GM_{*}/a^{2}\sim 70$ au, being $a\approx 11$ km s$^{-1}$ the sound speed in the ionized medium and adopting the stellar mass as $M_{*}\sim 10\ M_{\odot}$. Similarly, an ionized disk wind is expected to provide significant EM only at smaller radii from the star for those somewhat more massive stars explored in \citet{hollenbach1994}. The latter could replace the role of the jet discussed above and power a bipolar wide-angle wind driven by the pressure of the ionized gas \citep{parker1958} in a similar scenario proposed by \citet{keto2007}. Although a two-element scenario (HCHII region plus disk wind) is plausible because it matches the observational characteristics of G34.13-VLA1, it must be assessed through higher angular observations than those presented here.

%by using the~\citet{1985ApJ...288..618R}~\citep[with $A_{V}\sim 36$ mag estimated by][for this region]{2008MNRAS.391.1527P}, is in accordance with the photospheric contribution to the dust envelope at this band. 

%Furthermore, in agreement to \citet{keto2007}, the optical depth ($\tau$) of small hypercompact HII regions around massive stars is $\tau >1$ at our observed frequencies, which would be expected for a HII region trapped by a B2V star (for this star, with a mass   at $T_{e}\sim 10^{4}$ K).  

%On the other hand, if the excitation of the G34.13-VLA1 is occurring externally~\citep[see e.g][for a similar analysis in Sygnus OB2 region]{2019A&A...627A..58I}, its radio flux would come from ionized neutral material contained in the CRS provided that the responsible star delivers the ionizing photons ($N_{*}$) needed to maintain the flux seen in G34.13-VLA1~\citep{lugo2004}. 

\subsection{The G40.54 Region}

\subsubsection{Radio continuum}
\label{sec:result-G40-radio-continuum}

\begin{figure}
\begin{center}
\includegraphics[width=\columnwidth]{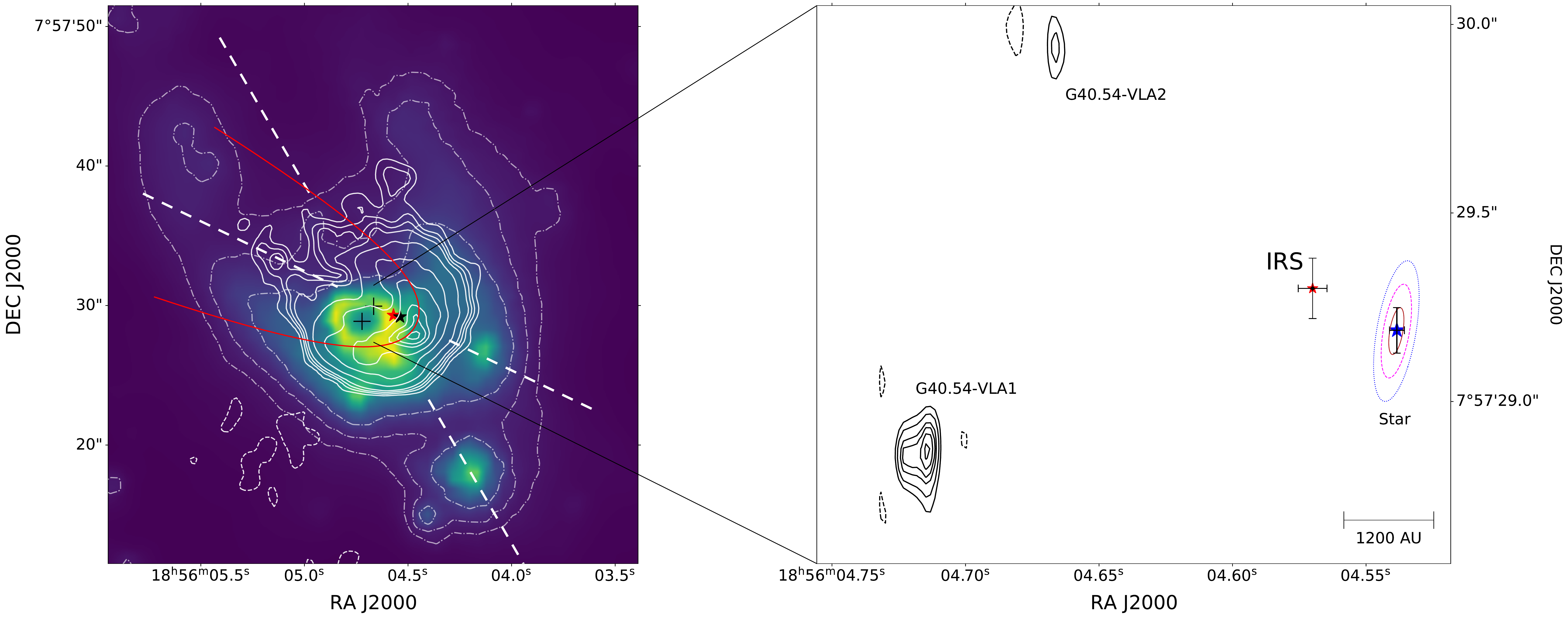}
\end{center}
\caption{\emph{Left panel:} 6 cm map (solid contours) and 4.5 $\mu$m GLIMPSE map (dashed contours) over-imposed on the same 4.5 $\mu$m map (color scale) of the G40.54 region. The solid contours are -3.0, 3.0, 5.0, 7.0, 10.0, 20.0, 40.0, 60.0, 80.0, 90.0, and 100.0 times the sigma rms noise (1$\sigma$ = 0.2 mJy beam$^{-1}$) while the dashed contours are 3.0, 5.0 and 10.0 times the rms noise of 1$\sigma\sim$12 MJy/sr. The curved red line shows the best adjustment of the~\citet{1996ApJ...459L..31W} bow-shock model, with the black ``star'' and the coincident white dashed line marking the best fitting position of the star and the position angle of the model (PA $\approx 116$ deg). The other white dashed line crossing nearly symmetrically the mid-IR contours denote the position angle of this emission ($\approx 153$ deg). The cross symbols mark the positions of the CRSs, and the red ``star'' marks the position of the IR sources found in the 2MASS and the UKIDSS catalogs. \emph{Right panel:} A zoom-in on the zone where the objects mentioned above lie. The contours show the 0.9 cm emission of G40.54-VLA1 and VLA2. This map was CLEANed with a weighting = ``BRIGGS'' and ROBUST = 0.0. The error bars on ``IRS'' and ``Star'' mark the 1$\sigma$ uncertainty in the position. The blue, red, and cyan dashed ellipses represent the 3, 2 and 1$\sigma$  uncertainty in the fitted position, respectively. For this map, contours are -4.0, 4.0, 5.0, 6.0, 6.5, 7.0, 7.5, 8.0, 9.0, 11.3 and 16 times the 1$\sigma$ noise level of 16.5 $\mu$Jy.}
\label{fig:adjusted-bow-shock_poss_star}
\end{figure}

In this region, we have detected two CRS, as shown in Figure.~\ref{fig:G40.54_ourcontourRFCsuper}, with the northern CRS being unresolved and the southern marginally resolved. These sources lie out of the bright cometary bow seen at 6 cm emission, with the southern CRS, G40.54-VLA1, appears to be marginally stretched in the east-west direction with two apparent lobes that turn up only when weighting ``Briggs'' is used to construct the map, such as seen in the right panel of {Figure}.~\ref{fig:adjusted-bow-shock_poss_star}. 
Given the solid detection of these (sub)structures in the maps across all the inspected uv ranges (see previous section) we consider them real instead of map artifacts. This shape tentatively resembles that of a proplyd-like morphology~\citep{1996ApJ...465..216H, raga2005} or a fragmented photo-evaporating clump~\citep{raga2005}.

According to the radio flux density, G40.54-VLA1 and VLA2 are ionized by $N_{*}\sim 10^{44}$ s$^{-1}$ photon rate. In an equilibrium situation when the exciting star is embedded in the CRS, similar to the explored for G34.13-VLA1, this flux would correspond to a B2V star or B3III type. However, this star could correspond to an earlier type if we consider that only a fraction of its total delivered photon rate is employed to ionize the CRS, such as resulting from a non equilibrium configuration with photon leakage. As seen in Fig. 2, G40.54-VLA1 has associated UKIDSS-GPS K-band emission that following Viveros et al. in prep. corresponds to a ZAMS B0 star. Hence, if G40.54-VLA1 harbors the exciting star of the region, only part of its radiation contributes to the observed free-free emission of this CRS, and the rest of UV photons would contribute to ionize the whole UCHII region. On the other hand, no IR counterparts are detected in G40.54-VLA2, implying the lack of associated stellar objects massive enough to be detected.    

%The observed radio emission represents only a fraction of the UV photon rate delivered by the star 

%. Likewise, for this type of star, the gravitational radius is $r_{g}\equiv GM_{*}/c^{2}\sim 77$ au, significantly smaller than the effective radius, ($R_{\mathrm{eff}}==\theta_{M}\theta_{m}$) of G40.54-VLA1 ($\approx 250\pm$ 77 au as reported in column 5 of the {Table}.~\ref{tab:physical-param}). Thus, for radius $r>r_{g}$, most of the ionized gas should escape from the gravitational force, so that the optically thin part of the CRS emission is not trapped but possibly outflowing as a result of its hydrodynamical pressure. As a result of that expansion, the UV opacity of the material is reduced by a Doppler shift, generating an underestimation in the counting of photons from the radio flux ($\sim 10^{44}$ s$^{-1}$). Thus, that non equilibrium situation suggests that the spectral type of the stellar object presumably associated to G40.54-VLA1 can be  earlier. Indeed, we found K-band magnitude associated to G40-VLA1 ($\approx 12$ mag), and following the same discussion of the previous section, (with $A_{V}\approx 37$ mag estimated from the color-magnitude diagrams (CMD) as shown in {Figure}.~\ref{fig:color_color_diagrams}), gives a ZAMS B0 star. In the case of G40.54-VLA2, its very lower limit in size ($R_{\mathrm{eff}}\sim 100$ au) and optical depth 

Another possibility is that G40.54-VLA1 and/or VLA2 are externally ionized by any massive star embedded in the region. In order to assess this possibility we revised the UKIDSS catalogue and found a central point infrared source (IRS) at the position of ($\alpha$, $\delta$)$_{\mathrm{J2000}}$ = 18$^{\rm h}$56$^{\rm m}$05.569$^{\rm s}$, +07$^{\circ}$57$'$29.30''. IRS is located 2.3 arcsec northwestwards of G40.54-VLA1 and 1.6 arcsec southwestwards of G40.54-VLA2 (see {Figure}.~\ref{fig:adjusted-bow-shock_poss_star}) and color-magnitude diagrams points to a main-sequence O6 star or earlier with a visual extinction of $\sim 37$ mag (Viveros et al. in prep.) for this source. 
Such a massive star is a plausible candidate to contribute significantly in ionizing G40.54-VLA1 and G40.54-VLA2 and, at same time, the whole G40.54 UCHII region. Following the same procedure as in Sect. 3.1.1 but for an O6 star, we estimate the photon flux rate impinging on G40.54-VLA1 and VLA2, which are shown in the table. Concerning the ionization of the overall UCHII region, the discrepancy with the spectral type of the exciting star derived from radio data (a B0 type) reflects that possible dust absorption, the porosity of the ionized gas, or dropping in opacity due to an expansion process of the gas in the UCHII region~\citep{2009ApJ...703.1352K} were not taken into account.

%This spectral type is significantly earlier than the estimated from radio observations (a B0-0.5V star)     

Nevertheless, the Lyman photon flux delivered by IRS shown in the table, despite being an upper limit, is significantly greater than the required to switch on G40.54-VLA1 and -VLA2, and only invoking an unlikely scenario with an ad hoc geometry and/or unexpected small size for the CRSs (that would imply a somewhat different nature than a free-free emitting object) would yield different results. We conclude then that IRS can ionise the nearby CRS G40.54-VLA1, G40.54-VLA2, and the whole G40.54 \uchii\ region.

\subsubsection{Bow-shock Models}
\label{sec:bow-shock model}

The cometary morphology seen at 6 cm is reminiscent of some interaction between the stellar activity and the ambient. Theoretical examples of bow shocks created by stellar winds have been widely explored, and in most of them, arc-shaped shells are produced, either in simulated images resulting from numerical models~\citep{1991ApJ...369..395M, 2015A&A...573A..10M} or analytical expressions~\citep{1990ApJ...353..570V, 1996ApJ...459L..31W, 1996ApJ...469..729C}. In our scenario, two of these models will be tested: the bow-shock model due to the stellar wind of a single moving star and the wind interaction model between two stars. We chose these models because they provide analytical expressions easily to analyze under the Montecarlo methods.

\emph{Single moving stellar wind}. \citet{1996ApJ...459L..31W} found the exact algebraic solution for the arc-shape shell formed by the wind interaction of a massive star with its molecular environment. More precisely, a star moving at a certain speed, $v_{*}$, concerning its surrounding homogeneous material of density, $\rho_{a}$, and losing mass at a rate, $\dot{M}_{w}$, produces an ionized shell shape that, projected in the plane of sky, is described by

\begin{equation}\label{eq:bow-shock-equation}
    R(\theta) = R_{0}\csc\theta\sqrt{3(1 - \theta\cot\theta)},
\end{equation}

\noindent where $R(\theta)$ is the distance between the star and the shell, at some opening-angle, $\theta$, seen from the star, with respect to the symmetry axis of the cometary morphology. The nearest distance, $R_{0}$, between the star and the bow-shock (the point on the arc where the ram and ambient medium pressure balance occurs) is given by:

\begin{equation}\label{eq:stand-off_distance}
 R_{0} = \sqrt{\frac{\dot{M}_{w}v_{w}}{4\pi\rho_{a} v_{*}^{2}}},    
\end{equation}

where $v_{w}$ is the wind velocity of the star, that for young massive stars, is about $\sim 10^{3}$ km s$^{-1}$. We fitted this model with a Markov Chain Monte Carlo method and obtained the following parameters: 

\begin{equation}
  \begin{split}
    \alpha_{*}=18^\mathrm{h}56^\mathrm{m}04.54^\mathrm{s}\\
    \delta_{*}=07^\mathrm{\circ}57^\mathrm{'}29.19^\mathrm{''}\\
    R_{0} = 2928.40_{-62.11}^{64.41}\ \text{au}\\
    PA = 115.64^{+2.75}_{-2.67}\ \text{deg}\ \text{N}\rightarrow \text{W}.
    \end{split}
\end{equation}\

where ($\alpha_{*}$,$\delta_{*}$) is the theoretical position of the star, $R_{0}$ is the ``stand-off'' distance defined above, and $PA$ is the position angle. To illustrate the accuracy of our results, we have plotted the most probably curve that adjusts the intensity profile of G40.54, such as it is shown in the left panel of~{Figure}.~\ref{fig:adjusted-bow-shock_poss_star}. In the right panel of this Figure, we show a ``zoom-in'' on the zone where the theoretical star is located. All the sources appearing in this panel are located far away beyond the 3 sigma contours (1200, 6000 and 5000 AU for IRS, G40.54-VLA1 and -VLA2, respectively, in projection). Thus, the positions of the CRSs and the IR source do not match the theoretical position found with the bow-shock model fitting. However, some positional discrepancies might be due to the constraints in the model, such as not accounting for inhomogeneities in the ambient molecular medium or the lack of angular resolution for describing the thin shell layer assumption of the model. Given that the IR source is significantly closer to the fitted position ($\sim0.3$ arcsec) with respect to the CRSs and in line with the results of the previous section, we favor IRS  as the star driving the bow-shock shape, which in turn is the most likely source exciting the \uchiir\ and associated CRSs.

%In~{Table}.~\ref{tab:distance_star_CRS} we report the projected angular distances between the IRS star and the CRSs G40.54-VLA1 and VLA2, and the respective UV photon rate impinging on the CRSs, $N_{A}$, which is $N_{*}\approx 2\times 10^{49}$ s$^{-1}$ for the CRSs in G40.54. Thus, according to the values of the table, it is very feasible that IRS provides the ionizing photon necessary to account for the radio emission seen in the CRSs G40.54-VLA1 and -VLA2.

With the value of $R_{0}$, and the expression in {Eq}.~\ref{eq:stand-off_distance}, we can estimate the velocity of the theoretical star with respect to its molecular environment. For this, we adopted characteristic mass loss rates of $\dot{M}_{w}\sim 6\times 10^{-7}$ $\mathrm{M}_{\odot}$ yr$^{-1}$, and wind speeds of $v_{w} =$ 2500 - 3000 km s$^{-1}$ \cuna\citep[both of these observed in O6 stars,][]{2003ApJ...599.1333S}. Typically, a molecular core has densities and mean molecular weights of $n_{\mathrm{H}_{s}}\sim 10^{4}$ and $\mu = 2.4$,  respectively. With this, we obtained a velocity of $V_{*} \sim 10$ km s$^{-1}$, higher than the average values found in star-forming regions\cuna\citep[e.g.][in the Orion nebula]{2017ApJ...834..139D} but still plausible\cuna\citep[e.g.][]{rodriguez2020}.

In the left panel of {Figure}.~\ref{fig:adjusted-bow-shock_poss_star}, it can be appreciated a similar morphology for the radio and mid-infrared emission, which implies a direct interaction between ionized gas and dust. However, the position angle of the IR emission is somewhat lower than that resulting from the fitting bow-shock model in the radio map ($\delta PA\approx 38$ deg). The latter suggests that a different dynamical process is responsible for shaping the dust content in the region; for instance, a champagne flow or ``blister'' process~\citep{tenorio-tagle1979, yorke1983}. In other words, a massive star, moving through a high-density medium, has break-out its natal molecular cloud core due to the ionization front expansion. By studying the relative motion, this hybrid model (bow-shock of a moving star plus a champagne flow) has already been inferred in the DR21 HII region \citet{2014A&A...563A..39I} of the ionized gas through H66-alpha radio-recombination line observations with respect to its ambient molecular gas. Despite this frame appearing to be highly favorable, high angular resolution radio-recombination lines and molecular observations would be necessary to assess this interpretation.

\emph{Two wind interaction model}. A two-wind interaction model was developed by ~\citet{1996ApJ...469..729C}. This model assumes that a bow-shock structure arises from the collision between isotropic winds of two stars. According to their results, the interacting profile surface of these winds should form a thin shell profile that can be described through the geometrical relation:

\begin{center}
\begin{equation}
    R = D \sin\theta_{1}\csc(\theta+\theta_{1}),
\end{equation}
\end{center}

where $R$ is the distance between the star internal to the arc and this structure,  $D$ is the separation between the two stars, and $\theta_{1}$ and $\theta$ are the angles subtended from the line joining the external and internal star with any given point in the arc, respectively. The hydrodynamical solution that relates these two angles with the momentum ratio of the stars, $\beta \equiv \dot{M}v/\dot{M}_{1}v_{1}$ %$\beta \equiv \dot{M}_{1}v_{1}/\dot{M}v$, 
is given by the expression:

\begin{equation}
\theta_{1}\csc\theta_{1} = 1 + \beta (\theta\cot\theta - 1).
\end{equation}
 
\citet{1996ApJ...469..729C} found an approximate solution to the above transcendental equation to express $\theta_{1}$ in terms of $\theta$ as

\begin{equation}
\theta_{1}\approx \left\{\frac{15}{2}\left[-1 + \sqrt{1 + \frac{4}{5}\beta (1 - \theta\cot\theta )}\right]\right\}^{1/2}
\end{equation}

In this case, $\beta<1$ implies that the internal star is less massive than the external one. Thus, according to this last relation, the higher the difference of mass or distance between the stars, the lower the value of $\beta$. The relation between $\beta$ and $D$ is given by

\begin{equation}
    R_{0} = \frac{\beta^{1/2}D}{1 + \beta^{1/2}},
\end{equation}

where $R_{0}$ is the ``stand-off'' distance between the internal star and the bow-shock head. In numerical terms, to guarantee the convergence of the fit, we fixed in separate trials some parameters of the model based on the IR data analyzed previously, namely the position and/or the spectral type of the two stars (that gives $D$, $\dot{M}$ and $v$). As possible positions for the internal star, we set those of the IRS source and the CRSs. For the external star, we adopted positions of several IR point sources of the 2MASS-PSC and UKIDSS-GPS catalogs that were visually located ahead of the bow-shock. The spectral type of some of them, including IRS, were derived from the IR analysis of Viveros et al. in prep. None of these combinations yielded satisfactory fits to the arc profile. Therefore, the model involving a single moving star analyzed above explains better the cometary morphology seen in G40.54.

\section{Discussion}
\label{sec:discussion}

The possibility that our discovered CRSs in the G34.13 and G40.54 \uchiirs have a photo-evaporative nature agrees with \cuna\citet{masque2017}. In that work, the authors claim that all \uchiirs must have at least one CRS associated due to the influence that CRSs have on the dynamical evolution of these regions. It is worth mentioning that our detected CRSs have faint emissions ($<1$ mJy) and only are detectable with deep enough observations, which explains why most of the CRSs in UCHII regions remained undetected in previous surveys. 

The massive stellar content of our \uchii regions is beyond doubt \citep[e.g.,][]{2012A&A...542A...3S}, but no clear association between CRSs and massive stars could be assessed in the previous section. Nevertheless, this implies that, either internally or externally, our CRSs are being exposed to a considerable amount of UV irradiance. Regardless of the stellar nature of CRSs, they likely possess a neutral part (i.e., a protostellar disk or neutral clump) that is being continuously ablated by the UV radiation, giving rise to the amounts of free-free emitting observed gas. Thus, in the following, we will investigate under this assumption the influence that the photo-evaporating activity of CRSs exerts on the dynamical evolution of the hosting UCHII region. More precisely, we shall probe if the CRSs are replenishing sufficient
ionized gas to the environment to maintain the UC H II region stage up to the time inferred statistically from the classical surveys \citep[$\sim10^6$yr,][]{wood-churchwell1989}. To do this, we first estimate the rate of mass loss of a CRS that, according to \cuna\citet{1978ppim.book.....S}, is:  

\begin{equation}\label{eq:mass_loss_CRS}
    \dot{M}_{\mathrm{crs}} = 2\pi R^{2} \frac{J_{0}\mu_{i}}{1 + \sqrt{1 + \frac{\alpha_{B}N_{*}R}{3\pi d^{2}C_{II}^{2}}}}, 
\end{equation}

where $R$ is the effective radius of the CRSs, $\alpha_{B}$ is the recombination coefficient ($\approx 3\times10^{-13}\ \mathrm{cm^{3}}\ \mathrm{s}^{-1}$), $C_{II}$ is the sound speed in the ionized medium ($\approx 11$ km s$^{-1}$), and $J_{0}=N_{*}/4\pi d^{2}$ is the diluted photon-flux from the ionizing star. In this expression $N_A=\pi R^{2} J_0$ and is delivered by a B0 star or a O6 star separated $d_\mathrm{CRS}$ from the CRS (see previous section), for G34.13 and G40.54, respectively. This assumes that only a fraction of the UV photons produced by the exciting star is employed in photo-ionizing the CRS, and the rest of them contributes in maintaining ionized the surrounding UCHII region. The effective radius $R$ is the most uncertain parameter because it is based on deconvolved sizes below our angular resolution, yielding only upper limits in most cases. However, we expect actual sizes of at least $\sim100$ AU \citep{dzib2013, masque2017}, otherwise smaller sizes would imply a too small free-free emitting region to be detected. With the parameters of Tables \ref{tab:obs-results} and \ref{tab:uchii-properties} we obtained, $\dot{M}_{\mathrm{crs}}\leq 1.4\times 10^{-6}$ $\mathrm{M}_{\odot}\mathrm{yr}^{-1}$ for CRS G34.13-VLA1, and  $\dot{M}_{\mathrm{CRS}}=2.7\times10^{-6}$ and $\leq 7.2\times 10^{-7}$ $\mathrm{M}_{\odot}\mathrm{yr}^{-1}$, for CRSs G40.54-VLA1 and -VLA2 respectively. 

Under our hypothesis, $\dot{M}_{\mathrm{crs}}$ contributes to filling up a volume with ionized mass $M$ that constitutes the "observed" UCHII region. Then, a typical replenishment time can be estimated by $\tau_\mathrm{rep}\approx M/\dot{M}$, being $M$ the total observed ionized mass of the region. For a hydrodynamic expansion, this occurs in roughly a crossing time (t$_\mathrm{ct}$). We do not follow the classical prescription of \citet{1978ppim.book.....S} to estimate $\tau_\mathrm{rep}$ because our situation differs in that a continuous replenishment of ionized gas occurs in the UCHII region. In the limiting case of $\tau_\mathrm{rep}\sim$t$_\mathrm{ct}$ the ionized gas flows outwards freely, producing, under ideal conditions, a density gradient. The observed UCHII region would correspond to the densest part of the flow, permanently observable as long as the CRS feeds it. The rest of the gas beyond the observed UCHII region would become too diluted to be detected by arrays. In a more realistic scenario supported by the morphologies analyzed in the previous section, however, some of the ionized gas is retained and accumulated in the observed UCHII region, giving $\tau_\mathrm{rep}$ values greater than t$_\mathrm{ct}$. The opposite case of having $\tau_\mathrm{rep}$ shorter than t$_\mathrm{ct}$ involves some acceleration mechanism pushing the material outwards the UCHII region, such as stellar winds or radiation pressure.       
 
%In general, the sound crossing time of the CRS must be extremely small given their typical compact size \citep[a few hundreds of AU,][]{masque2017}. Figure \ref{fig:replenishing-crs} compares this time with the time need to replenish the CRS volume for the CRS G40.54-VLA1, the only one for which an estimation of its deconvolved size could be obtained. In this calculation, we used the mass loss rate for this CRS. The intersection between dashed and solid lines is the time need for the CRS to supply the ionized material estimated for the whole UCHII. Note that the errors are large but consistent with a free expansion replenishment, discarding other mechanisms with additional powering like winds that would need shorter times (e.g. a few years).

\begin{figure}
\begin{center}
    \includegraphics[width=0.45\textwidth]{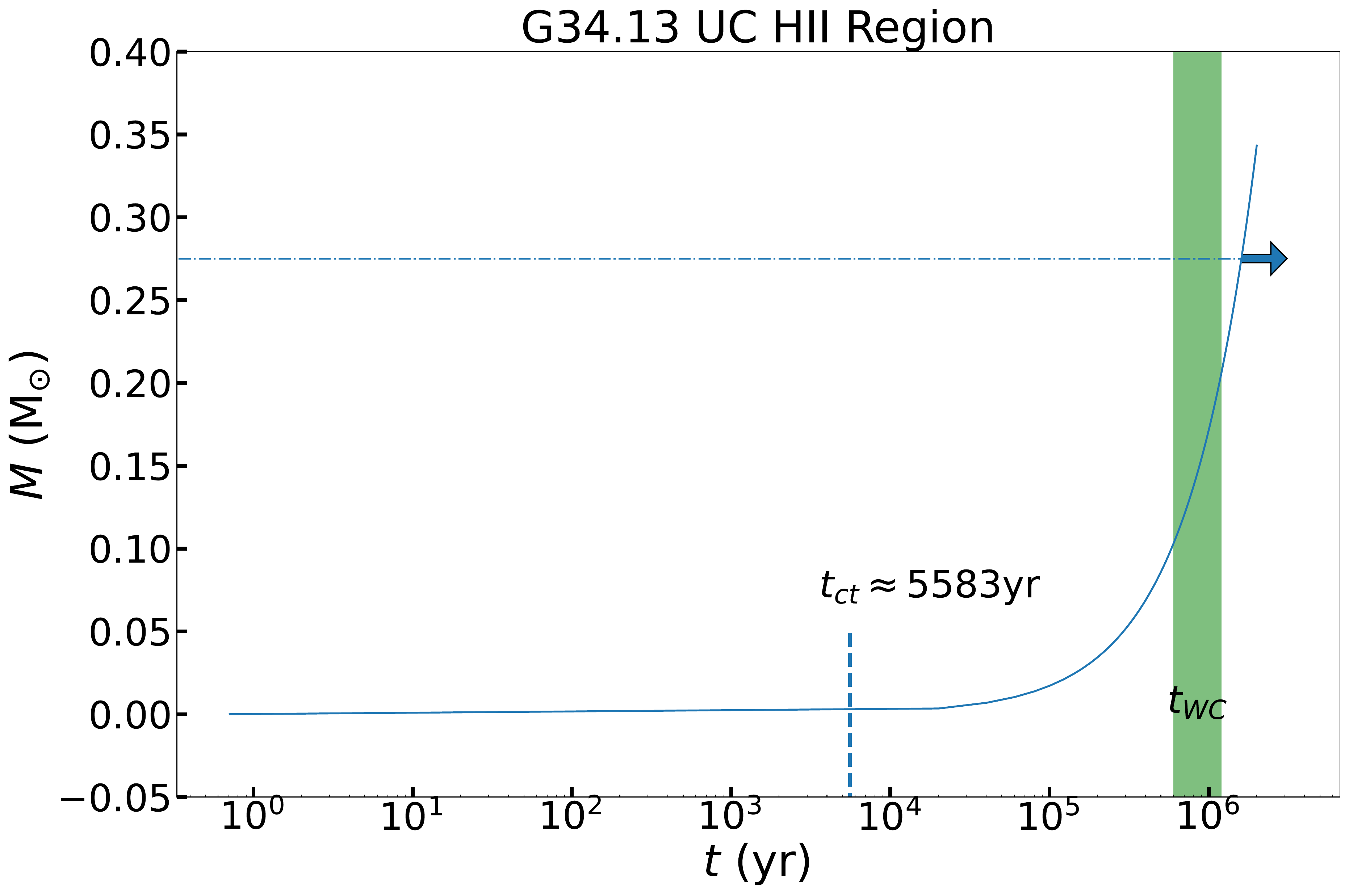}
    \includegraphics[width=0.45\textwidth]{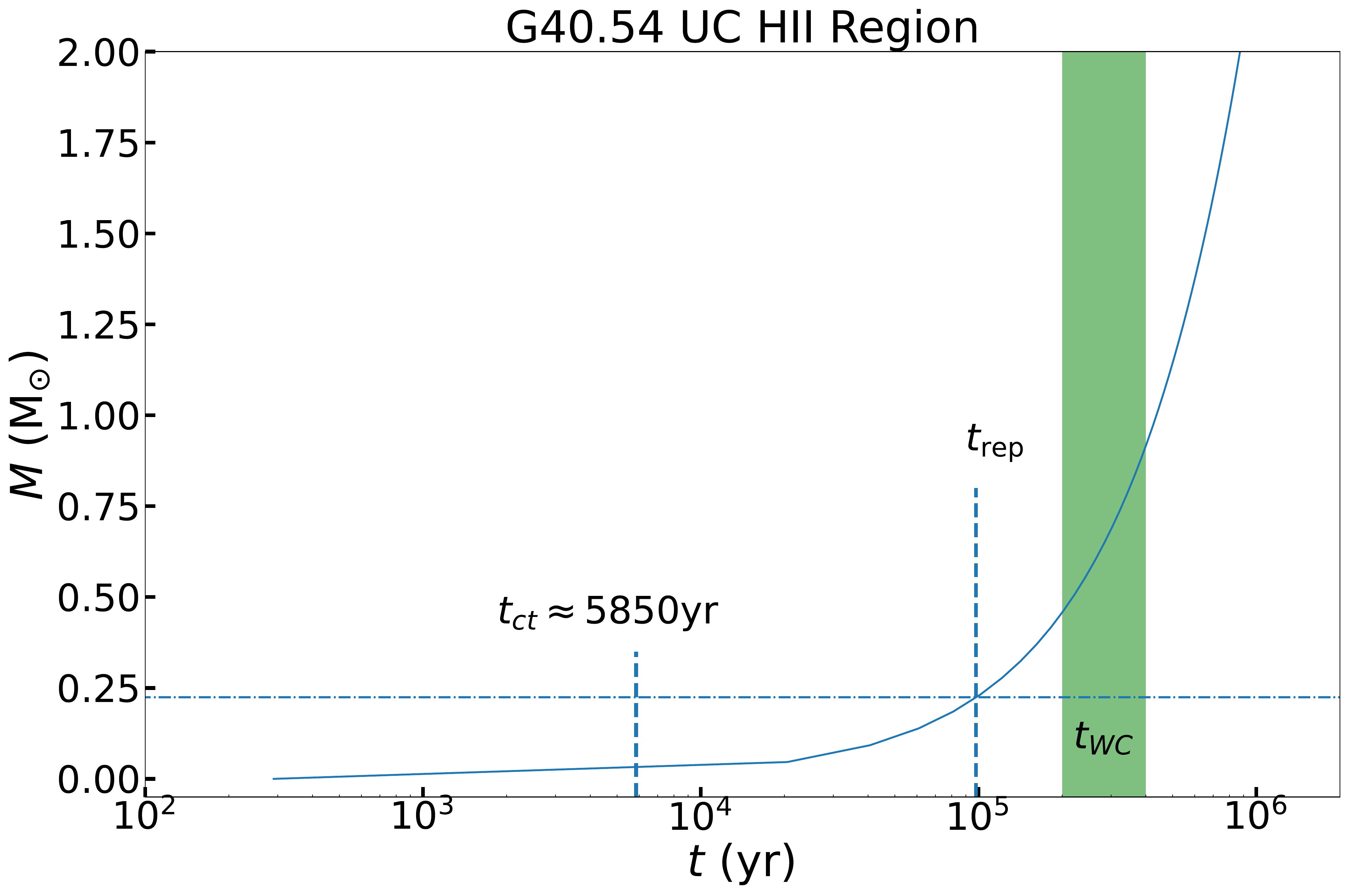}
\end{center}
\caption{Total mass loss (=$\dot{M}_{crs}$t) vs. time for the CRSs for G34.13 (left panel) and G40.54 (right panel). The continuum blue line is $\dot{M}_{crs}$ while the horizontal ``dashed'' line marks the current total ionized hydrogen mass of the UCHII region (0.28 M$_{\odot}$ and 0.22 M$_{\odot}$ for G34.13 and G40.54 respectively). The vertical green ``shaded'' band depicts the range of lifetimes for UCHII regions estimated statistically in \citet{wood-churchwell1989}. The replenishment (t$_{rep}$) and crossing times (t$_{ct}$) are shown with vertical dashed lines. Due to G34.13 is not spatially resolved, the value of t$_{\rm rep}$ can not be determined. Nevertheless, the starting of the arrow (top-right) indicates the lower limit for t$_{\rm rep}$ (1.6$\times$10$^6$). 
}
\label{fig:replenising-uchiirs}
\end{figure}

In\cuna{Figure}.\cuna\ref{fig:replenising-uchiirs} we show the plot $M$ vs $t$ for the G34.13 and G40.54 UCHII regions. The two horizontal dashed lines indicates the 'observed' mass of our UCHII regions at present. They were estimated using VLA L-band data taken at B configuration for G34.13 \citep{2005AJ....130..586W} and at D configuration for G40.54 \citep{1998AJ....115.1693C}, and following ~\citet{1967ApJ...147..471M}, giving 0.28 and 0.22 $\mathrm{M}_{\odot}$ for G34.13 and G40.54, respectively. In these plots we show t$_\mathrm{ct}$ ($\sim5 \times 10^3$ yr ) and the statistical $t_\mathrm{WCH}$ ($\sim10^6$yr) time estimated by \citet{wood-churchwell1989} in terms of an uncertainty range, for comparison. The replenishment time $\tau_\mathrm{rep}$ is given by the intersection between the solid curve representing $\dot{M}_{\mathrm{crs}}$ and the horizontal dashed line.       

For G34.13, we only can derive a lower limit of $\tau_\mathrm{rep}$ because G34.13-VLA1 is spatially unresolved under our observations. Nevertheless, this limit ($>2 \times 10^6$ yr) is greater than the dynamic time t$_\mathrm{ct}$ meaning that, in spite of the mass loss caused by the outflow present in the region (see Sect. 3.1.2), most of the material remains trapped perhaps by the structures seen in IR \citep[e.g. a torus-like][]{2009MNRAS.393..354P}. However, $\tau_\mathrm{rep}$ is larger than the statistical time $t_\mathrm{WCH}$ that implies a replenishment time longer than the inferred from observations (or too low $\dot{M}_{\mathrm{crs}}$ to replenish the region efficiently). Therefore, the replenishing role assumed for G34.13-VLA1 seems unlikely. 

For G40.54, we took into account the mass contribution of G40.54-VLA1 and -VLA2, namely, $\dot{M}_{tot} = \dot{M}_{\mathrm{CRS1}} + \dot{M}_{\mathrm{CRS2}}$. From the plot, we can see that $\tau_\mathrm{rep}$ falls between t$_\mathrm{ct}$ and $t_\mathrm{WCH}$, enabling the replenishing possibility in G40.54 and matching well with some retention of material in the region as a natural consequence of its accumulation in the bow shock modelled in previous section. Moreover, as discussed in that section, the surrounding neutral cloud may help confine the region as inferred from IR maps.  

The dynamical evolution of UCHII regions is a complex issue, and in this paper, we aimed to provide plausible explanations for the observed long lifetime and morphology of these regions, aiding future investigations on this subject. While our findings in G40.54 make the 'replenishing' possibility viable, additional elements must be included that, especially in  G34.13, must be dominant. For example, the CRSs could have been fragmented out from originally larger structures such as pillars or elephant trunks \citep{dzib2014, cortesrangel2020}, implying higher 'replenishment' during longer times, though this process depends on the strength of the energetic radiation field impinging on such structures \citep{raga2005}. This is in perfect agreement with the multiple CRSs seen in G34.13 or the G35.20 UCHII region \citep{masque2021}. Furthermore, some IR sources spatially associated with the UCHII regions could be an evolutionary product of CRSs, with their neutral part already consumed by the energetic radiation, remaining only \textit{naked} YSOs. Characterizing the neutral part of the CRSs with high-quality millimetric data like those provided by the ALMA telescope is paramount to accurately estimating CRS lifetimes and assessing this proposed evolutionary picture. 

\section{Conclusions}
\label{sec:conclusion}

We carried out deep VLA observations with a high angular resolution of the G34.13 and G40.54 \uchii regions to explore their most compact components. From these observations and the subsequent analysis of the data, we achieved the following conclusions:     

\begin{itemize}
    
     \item We found three CRSs (sizes of $\sim100$ AU) associated with the G34.13 (G34.13-VLA1) and G40.54 (G40.54-VLA1 and G40.54-VLA2) \uchiirs. The detection of these CRSs in only two explored UCHII regions is following \cuna\citet{masque2017}, claiming that the presence of CRS in UCHII regions is a common phenomenon.  
     
     \item The radio flux density of our CRSs is compatible with internal ionization by a B3 type star, significantly less massive than the stars needed to excite their whole hosting UCHII regions. The latter opens the possibility that CRSs does not necessarily host massive stars and, instead, they may be associated to close lower mass companions.        
     
    \item The derived high emission measures of our CRS ($\sim 10^{7}$ cm$^{-6}$pc) discard the presence of single isolated radio jets in them. Otherwise, other scenarios like trapped HCHII regions and disk winds must be dominant in producing the free-free emission of the CRS. For the particular case of G34.13-VLA1, the large-scale outflow features observed previously over the G34.13 region can be explained by a photo-evaporated disk wind driven by the pressure of ionized gas instead of being powered by a \textit{classical} radio jet in the CRS.     
    \item The cometary morphology of the UCHII region G40.54 could result from the activity of a massive star moving at 10 km s$^{-1}$ across the region, aided by the density structure of the ambient cloud. This star is likely associated with an IR source, nearby to but not spatially coincident with G40.54-VLA1 nor -VLA2, and it is the most plausible candidate in photo evaporating the CRSs in the case of external irradiation. The proposed scenario probes the role of massive stars shaping the fresh ionized material provided by CRSs, generating some of the morphologies observed for UCHII regions.  
    
    \item Under the photo evaporating scenario, in some cases (e.g., G40.54), the CRSs can supply material to the UCHII region at a sufficiently high rate to become observable for a significantly longer time. In other cases (e.g., G34.13), additional mechanisms driving the evolution of UCHII regions should be invoked. Therefore, there is no unique model (i.e., replenishing or confining) to address this subject, and a combination of them must be employed in each case.

\end{itemize}

%%%%%%%%%%%%%%%%%%%%%%%%%%%%%%%%%%%%%%%%%%%%%%%%%%%%%%%%

\begin{ack}

The authors thank the anonymous referee for the helpful and constructive comments that improve the manuscript. EdelaF thanks PRODEP-SEP through Cuerpo acad\'emico UDG-CA-499, Coordinaci\'on General Acad\'emica y de Innovaci\'on (CGAI-UDG), authorities of CUCEI, and both, academic and administrative staff of the Institute for Cosmic Ray Research (ICRR) of the University of Tokyo (UTokyo) for financial support and management during his Sabbatical stay at the ICRR-UTokyo in 2021. MATH acknowledges support from Direcci\'on de Apoyo a la Investigaci\'on y al Posgrado (Universidad de Guanajuato), grant 048/2022.

\end{ack}

\section*{Data availability}

The data underlying this article are available in NRAO Science Data Archive [Karl Jansky VLA Database] at https://archive.nrao.edu/archive/archiveproject.jsp, and can be accessed with the respective Project (Proposal) Code: AK423

\bibliographystyle{dcu}

\bibliography{article_CRS}

\appendix
\section{Physical Parameters}\label{apx:physical-params}

The physical parameters of compact radio sources were derived by using the expressions reported by~\citet{2020MNRAS.497.4436D} in their appendix. The main-beam brightness temperature was calculated with the \textit{Reyleigh-Jean} approximation of a black body\

\begin{equation}
\left[\frac{T_{\mathrm{B}}}{\mathrm{K}}\right] = 1.62 \left[\frac{S_{0.9\mathrm{cm}}}{\mathrm{mJy}}\right]\left[\frac{\theta_{\mathrm{M}}}{\mathrm{arcsec}}\right]^{-1}\left[\frac{\theta_{\mathrm{m}}}{\mathrm{arcsec}}\right]^{-1},  
\end{equation}\

\noindent where we have used the solid angle expression: $\Omega \approx 1/2\pi\theta_{M}\theta_{m}$.

The ratio between the brightness temperature and the electron temperature differs by a factor of $1 - e^{-\tau_{\nu}}$, where $\tau_{\nu}$ is the optical depth. From this, we could calculate the opacity by assuming a electron temperature of $10^{4}$ K.

From~\citet{1967ApJ...147..471M}, we have used the radio frequencies expression to calculate the emission measure, EM, as follows\

\begin{equation}
\left[\frac{EM}{\mathrm{cm}^{-6}\mathrm{pc}}\right] = \left[\frac{T_{e}}{\mathrm{K}}\right]^{1.35}\left[\frac{\nu}{\mathrm{GHz}}\right]^{2.1}\tau_{\nu}.
\end{equation}\

Following the definition of EM=$n_{e}^2dl$, it was possible to estimate the electron density $n_{e}$, by adopting the line-of-sight depth, $l$, as the linear dimension of the source, with the expression\

\begin{equation}
    \left[\frac{n_{e}}{\mathrm{cm^{-3}}}\right] = 454.08\left[\frac{EM}{\mathrm{cm}^{-6}\mathrm{pc}}\right]^{0.5}\left[\frac{\theta_{\mathrm{eq}}}{\mathrm{arcsec}}\right]^{-0.5}\left[\frac{D}{\mathrm{pc}}\right]^{-0.5},
\end{equation}\

\noindent where $\theta_{\mathrm{eq}}$ = $\sqrt{\theta_{M}\theta_{m}}$.

The ionizing photons rate can be calculated by assuming that a balance between recombination and ionization rates in the \hiirs has been established, namely~\citep{1939ApJ....89..526S}\

\begin{equation}
\left[\frac{\dot{N}}{s^{-1}}\right] = 1.75\times 10^{39}\left[\frac{\theta_{\mathrm{eq}}}{\mathrm{arcsec}}\right]^{3}\left[\frac{n_e}{\mathrm{cm}^{-3}}\right]^{2}\left[\frac{D}{\mathrm{pc}}\right]^{3}\alpha.
\end{equation}\

We have taken the common value for the recombination coefficient, $\alpha$ = $3\times 10^{-13} \mathrm{cm}^{-3}\mathrm{s}^{-1}$~\citep{1987MNRAS.224..801H}.

\newpage

%======== FIGURES =======

%\begin{figure*}
%\begin{center}
%   \includegraphics[width=0.45\textwidth]{figs/dynamic_gas_CRS1_G40.pdf}
%\end{center}
%\caption{Logarithmic representation of the Mass loss vs time function at CRS scale for G40.54-VLA1 (blue continuum line). As representation of the uncertainty of the mass function, we have depicted, with black dashed lines, $M = (\dot{M}+\Delta\dot{M})\times t$ and $M = (\dot{M}-\Delta\dot{M})\times t$, where $\Delta\dot{M_{CRS}}$ is the uncertainty in the mass loss as calculated with {Eq}.~\ref{eq:mass_loss_CRS} taking into account the size error in {Table}.~\ref{tab:obs-results}. The horizontal orange error bar is the uncertainty in the time that the material fills the CRS volume with the photo-evaporated gas expanding freely, $t_{f,c}\approx R/C_{II}$.}
%\label{fig:replenishing-crs}
%\end{figure*}

% ===== TABLES =====

\begin{table*}
\caption{\label{tab:obs_outline}Outline of the observational parameters.}
\centering
\begin{tabular}{|p{5.0cm}|c|}
 \hline
 \hline
 Parameter & Value\\
 \hline
Center Frequency (GHz) . . . . . . . . . .   &  32.936 (K$_{\mathrm{a}}$-Band)\\
Bandwidth (GHz) . . . . . . . . . . . . . .    & 8.0\\
Primary beam . . .  . . . . . . . . . . . . .  & $\sim$ 1.25$'$ \\
Synthesized beam . . . . . . . . . . . .   &  $\sim$0.25$''$ \\
Largest imaged structure . . . . . . . . .   & $\sim$ 1.85$''$ - 0.5$''$ \\
Typical rms in image ($\mu$Jy beam$^{-1}$) . . & 19 \\
Theoretical RMS ($\mu$Jy beam$^{-1}$). . . . .  & 10 \\ 
Primary flux density calibrator, 3C 286 (Jy). &  0.7 \\ 
 \hline
\end{tabular}
\end{table*}

\begin{table*}
\caption{Parameters of the Compact Radio Sources obtained from Gaussian fits.\label{tab:obs-results}}
\begin{center}
\resizebox{0.9\textwidth}{1.5cm}{
\begin{tabular}{cccccccc}
\hline
\hline
& \multicolumn{2}{c}{Coordinates}  &   $\theta_M$  x  $\theta_m$  ;  P.A. $^{\mathrm{a}}$ & $S_\mathrm{0.9cm}^{\mathrm{b}}$  & $I^{peak, \mathrm{c}}_\mathrm{0.9cm}$& RMS$^{\mathrm{d}}$& CRS\\
Region   &  $\alpha$~(J2000) & $\delta$~(J2000) &  ($mas  \times  mas$  ;   $^\circ$) &  (mJy)  &  (mJy bm$^{-1}$) & ($\mu$Jy bm$^{-1}$) & Name    \\
\hline\noalign{\smallskip}
G40.54 & $18^\mathrm{h}56^\mathrm{m}4\rlap{.}^\mathrm{s}717$ & $+07^{\circ}57'28\rlap{.}''84$ &  $ (317 \pm 111)  \times (142 \pm 70) ; 11 \pm 33 $ & $ 0.57 \pm 0.14$  & $ 0.24 \pm 0.04$ & 21 & G40.54-VLA1\\\noalign{\smallskip}
%G40.54 & $18^\mathrm{h}56^\mathrm{m}4\rlap{.}^\mathrm{s}722$ & $+07^{\circ}57'28\rlap{.}''87$ &  $ 195 \pm 58  \times 137 \pm 29  ;  171 \pm 24$  & $ 0.15 \pm 0.06$ & $ 0.13 \pm 0.03$ & 1.8 $\times$ 10$^{-5}$ & G40.54-VLA1-East\\
G40.54 & $18^\mathrm{h}56^\mathrm{m}4\rlap{.}^\mathrm{s}666$ & $+07^{\circ}57'29\rlap{.}''95$ &  $\leq (225 \pm 69)  \times (116 \pm 19) ;  7.8 \pm 7.3 $  & $ 0.12 \pm 0.05$ & $ 0.11 \pm 0.02$ & 21 &G40.54-VLA2\\
\noalign{\smallskip}
G34.13 & $18^\mathrm{h}51^\mathrm{m}57\rlap{.}^\mathrm{s}160$ & $+01^{\circ}17'00\rlap{.}''41$ &   $\leq (177 \pm 8)  \times (151 \pm 6);  17 \pm 10$    & $ 0.46 \pm 0.03$  & $ 0.44 \pm 0.02$& 15 & G34.13-VLA1\\ 
\hline
\end{tabular}}
\end{center}
$^\mathrm{a}$Angular size of the sources.\\
$^\mathrm{b}$Integrated flux density.\\
$^\mathrm{c}$Intensity peak.\\
$^\mathrm{d}$RMS averaged over the central part of the maps.
\end{table*}

\begin{table*}
\caption{Derived physical parameters from set one of equations.\label{tab:physical-param}}
\begin{center}
%\vspace{0.2cm}
%\scriptsize
\begin{tabular}{cccccccc}
\hline
\hline
 & $T_{b}$ $^{\mathrm{a}}$  &  $\tau$ $^{\mathrm{b}}$  &  EM $^{\mathrm{c}}$ & $n_{e}$ $^{\mathrm{d}}$ & Size  & $\log N_{CRS}$ $^{\mathrm{e}}$ & $M_{\mathrm{H~II}}^{f}$ \\
Fuente   &  K & (10$^{-3}$) &  (10$^{7}$cm$^{-6}$pc) &  (10$^{4}$cm$^{-3}$)  & au & (s$^{-1}$) & $10^{-6}\mathrm{M}_{\odot}$    \\
\hline\noalign{\smallskip}
%G40.54-VLA1-West	&	10.38$\pm$	&	1.04$\pm$	&	4.87	&	4.83	&	490	&	43.9	&	5.21	\\
G34.13-VLA1	&	$\geq$82.13	&	$\geq$0.008	&	$\geq$4	&	$\geq$18	&	$\leq$389	&	44.30	&	$\leq$3.23	\\
G40.54-VLA1	&	20.64$\pm$13.4	&	2$\pm$1.6	&	1$\pm$0.05	&	6.3$\pm$1.1	&	509$\pm$154 	&	44.37	&	18$\pm$1	\\
G40.54-VLA2	&	$\geq$22.0	&	$\geq$2	&	$\geq$1	&	$\geq$10	&	$\leq$225	&	43.69	&	$\leq$1	\\

\hline
\end{tabular}
\end{center}
$^\mathrm{a}$ Brightness temperature  \\
$^\mathrm{b}$ Optical depth\\
$^\mathrm{c}$ Emission measure \\
$^\mathrm{d}$ Electron density \\
$^\mathrm{e}$ Lyman photon flux needed to ionize the CRS\\
$^\mathrm{f}$ Mass of ionized hydrogen.\\

\end{table*}

\begin{table}
\caption{Lyman photon flux interaction parameters between the ionizing star and the CRSs \label{tab:distance_star_CRS}}
\begin{center}
 \begin{tabular}{c c c c} 
 \hline\hline
 RFC & $d_{\mathrm{CRS}}$ (au)$^{a}$ & $\log N_{A}$ (s$^{-1}$)$^{b}$ & $\log N_{\mathrm{CRS}}$ (s$^{-1}$)$^{c}$ \\ 
 \hline

 G34.13-VLA1 & 35000$^{d}$ & $\sim$ 45 & 44.30$^{d}$ \\ 
% \hline
 G40.54-VLA1 & 4000 & $\sim$ 46 & 44.37 \\ 
 G40.54-VLA2 & 3800 & $\sim$ 45 & 43.69 \\
% G40.54-VLA1 & 1.9 & $\sim$ 44.24 & 44.47 \\ 
% G40.54-VLA2 & 1.8 & $\sim$ 43.40 & 43.87 \\ 
 \hline\hline
\end{tabular}
\end{center}
$^{a}$ Projected distance between the CRS and IRS \\
$^{b}$ $\log$ of ionizing photon rate arriving to the CRS as calculated with expression {Eq}.\ref{eq:dilution}.\\
$^{c}$ $\log$ of ionizing photon rate as calculated from the radio flux density (see Table~\ref{tab:physical-param})\\
$^{d}$ The values correspond to the UKIDSS-GPS source detected in the G34 region closest to G34.13-VLA1, and assumes a B0.5V spectral type (see text).\\

\end{table}

\begin{table}
\caption{Review of the principal parameters of the \uchiirs. \label{tab:uchii-properties}}
\begin{center}
 \begin{tabular}{c c c} 
 \hline\hline
 Parameters & G34.13 & G40.54  \\
 \hline
$\theta_{\mathrm{maj}}^\mathrm{a}$ (arcsec)& 11.2 & 16.3  \\ 
$\theta_{\mathrm{min}}^\mathrm{a}$ (arcsec) & 10.5 & 11.0 \\
$S_{\nu}^\mathrm{a}$  (mJy)& 575.0 & 460.0 \\
$D^\mathrm{b}$ (kpc) & 2.47 & 1.91\\
$\nu$ (GHz) & 1.4 & 1.4\\
VLA-config & B & D\\
% G40.54-VLA1 & 1.9 & $\sim$ 44.24 & 44.47 \\ 
% G40.54-VLA2 & 1.8 & $\sim$ 43.40 & 43.87 \\ 
 \hline\hline
\end{tabular}
\end{center}
$^{a}$ Angular size and integrated flux density extracted from \citet{2005AJ....130..586W} and \citet{1998AJ....115.1693C} for G34.13 and G40.54, respectively.\\
$^{b}$ Distance to the regions taken from\cuna\citet{2015MNRAS.451.3089T} of their identified parental infrared dark cloud (IRDC) distance. For G40.54, the Bayesian near-distance result of\cuna\citet{reid2016} was adopted.\\
\end{table}

%\begin{longtable}{cllcccccc}
%\caption{Spectral Energy Distribution Results}
%\label{tab:tab1}
%\hline 
%Frequency  & Flux & Flux\_err & major\_axis & major\_axis\_err & minor\_axis & minor\_axis\_err & PA & PA\_err  \\
% (GHz)  & (Jy) & (Jy) & (mas) & (mas) & (mas) & (mas) & ($^{\circ}$) & ($^{\circ}$) \\ 
%\endfirsthead
%\hline
%\endhead
%\hline
%\endfoot
%\hline
%\endlastfoot
%\hline
%43.34000  & 0.04707 & 0.00376 & 99.1 & 4.5 & 27.8 & 1.5 & 176.7 & 1.1 \\
%22.45995  & 0.02481 & 0.00284 & 159.4 & 11.4 & 46.6 & 7.5 & 2.5 & 2.8 \\
%14.93962  & 0.01729 & 0.00247 & 234.0 & 8.5 & 40.5 & 12.1 & 178.3 & 1.5 \\
%8.459633  & 0.01006 & 0.00147 & 528.3 & 10.7 & 256.2 &  2.9 & 7.7 & 0.59 \\
%4.860100  & 0.00787 & 0.00093 & 582.0 & 25.0 & 125.0 & 39.0 & 178.9 & 2.2 \\
%\end{longtable}

\end{document}